\documentclass[structabstract]{aa}
\usepackage{graphicx}
\usepackage{amsmath,epsfig}
\usepackage{txfonts}
\begin{document}
   \title{The self-cohering  tied-array}

   \author{P. Fridman
          \inst{}
          }

   \institute{ASTRON, Oudehoogeveensedijk 4,  Dwingeloo, 7991PD, The Netherlands\\
              \email{fridman@astron.nl}
             }

   \date{}

% \abstract{}{}{}{}{}
% 5 {} token are mandatory

 % \abstract{}{}{}{}{}
% 5 {} token are mandatory

  \abstract
   {Large radio astronomy multi-element interferometers  are frequently used as  single dishes  in a tied-array mode  when  signals from separate antennas
are added. Phase shifts arising during wave propagation through a turbulent atmosphere can significantly reduce the effective area of an equivalent single dish.
   }
   { I aim to give estimates of  the impact of the ionosphere and troposphere  on the effectiveness of a radio interferometer working in  tied-array mode.
   }
   {Statistical estimates of the effective area are calculated and the power-law of  turbulent atmosphere irregularities has been used.
   A simple method of  tied-array calibration  using optimization techniques is proposed.}
   { The impact of  phase errors on the effectiveness of tied-arrays are given for low and high frequencies. Computer simulations
    demonstrate the efficacy of the proposed calibration algorithm.}
   {}


   \maketitle

\section{Introduction}
Large radio astronomy multi-element interferometers  (VLA, WSRT) are frequently used in the tied-array mode where  signals from separate antennas
are added (Thompson et al. 2001, ch. 9.9). The output sum signal  can be used in  VLBI,  pulsar  and transients  observations (Cordes 2004), SETI signals detection and the direct-to-Earth (DtE) reception of signals from cosmic apparatus (Jones 2004). In all these cases a radio interferometer works as a single-dish antenna
with one output. Partial signals from antennas are properly phased to collect emission from a point-like radio source in the sky and track it during its siderial movement. Standard calibration procedure using a correlator is employed to provide the necessary phase corrections for each individual antenna.  Random phase perturbations such as   phase shifts arising during wave propagation through the turbulent atmosphere can occur in  the course of such observations. These phase errors reduce the total effective area of the tied-array and must be compensated for in {\it real time}.  Although it is possible to store baseband
data for processing after  observations have taken place, the amount of data to be stored places a firm limit to the number of antennas that can be used in this manner.\\
New large scale projects such as SKA and LOFAR will  also be operating  in  tied-array mode. The impact of  ionospheric  and tropospheric  phase errors on the tied-array is calculated in this paper. A simple method of correcting  these errors using the output signal of the tied-array is also proposed here.

\section{Tied-array with random phase errors}
Voltage  produced by the source  at the output of the planar tied-array is
\begin{equation}
E({\bf s})=\exp (j2\pi ft)\sum_{n=1}^{n_{a}}a_{s,n}\exp [-j(kr_{s,n}+\delta _{n})]
\end{equation}
where  ${\bf s}$ is the source vector, $k=2\pi/\lambda$, $\lambda$ is the wavelength, $f$ is the frequency,  $a_{s,n} $- is the signal amplitude at n-th array element, $r_{s,n}$ -is the distance between the source and n-th array element, $n_{a}$ is the number of antennas in the array, $\delta _{n}$ is the instrumental phase shift. The distance $r_{s,n}$ can be represented as the module of the vector difference ${\bf s} -{\bf p}_{n}$, where ${\bf p}_{n}$ is the position vector of n-th array element:
\begin{eqnarray}
r_{s,n}=\sqrt{\left| {\bf s}-{\bf p}_{n}\right| ^{2}}=((s_{x}-p_{n,x})^{2}+(s_{y}-p_{n,y})^{2}+s_{z}^{2})^{1/2}= \nonumber\\
(r_{s}^{2}-2p_{n,x}s_{x}-2p_{n,y}s_{y}+p_{n,x}^{2}+p_{n,y}^{2})^{1/2}\approx r_{s}-\frac{p_{n,x}s_{x}+p_{n,y}s_{y}}{r_{s}}=\nonumber\\
  r_{s}-p_{n,x}\cos (\alpha _{s,x})-p_{n,y}\cos (\alpha _{s,y})=r_{s}-{\bf p}_{n}{\bf u}_{s},
\end{eqnarray}
$s_{x},s_{y},s_{z}$ are the components of vector ${\bf s}$  and $p_{n,x},p_{n,y}$ are the components of vector ${\bf p}_{n}$, ${\bf u}_{s}=[\cos (\alpha _{s,x}),\cos (\alpha _{s,y})]$.\\
With these new notations (1) can be rewritten as:
\begin{equation}
E({\bf u}_{s})=\exp (j2\pi ft-jkr_{s})\sum_{n=1}^{n_{a}}a_{{\bf u}_{s},n}\exp (-j[k{\bf p}_{n}{\bf u}_{s}+\delta _{n}]).
\end{equation}
An array is directed at ${\bf s}$ when these conditions are satisfied:
\begin{equation}
-k{\bf p}_{n}{\bf u}_{s}+\delta _{n}=0, n=1...n_{a}.
\end{equation}

The manifold of signals received in other directions (${\bf u} \neq {\bf u}_{s}$) forms the field pattern:
\begin{equation}
E({\bf u},{\bf u}_{s})=\exp [j2\pi ft-jk(r_{s}-r_{u})]
\sum_{n=1}^{n_{a}}a_{{\bf u},n}\exp [-jk{\bf p}_{n}({\bf u}_{s}-{\bf u})].
\end{equation}
The power pattern is the expected value of the product
\begin{eqnarray}
P({\bf u},{\bf u}_{s})=<E({\bf u},{\bf u}_{s})\overline{E({\bf u},{\bf u}_{s})}>=\nonumber\\
<\sum_{n=1}^{n_{a}}\sum_{m=1}^{n_{a}}\{a_{{\bf u},n}a_{{\bf u},m}\exp [-jk({\bf p}_{n}-{\bf p}_{m})({\bf u}_{s}-{\bf u)]\}>}
\end{eqnarray}
where $(\overline)$ denotes a complex conjugate and $< >$ denotes a time average.\\
In the absence of phase errors and in the direction of the radio source, i.e., for ${\bf u}={\bf u}_{s},a_{{\bf u},n}=a,P({\bf u}_{s},{\bf u}_{s})=
 n_{a}^{2}a^{2}$. The dc component proportional to the sum  of the system noise power at each antenna is omitted here and considered to be a constant value and therefore not relevant. \\
In the presence of phase errors produced by, for example, the  atmosphere, there are additional phase terms $\delta_{n,atm}\not=0$  in (6) and the power received in the direction ${\bf u}_s$ is:
\begin{eqnarray}
P_{\delta_{atm} }({\bf u}_{s},{\bf u}_{s})=\nonumber\\
a^{2}\sum_{n=1}^{n_{a}}\sum_{m=1}^{n_{a}}<\exp [j(\delta _{n,atm}-\delta _{m,atm})>=\nonumber\\
a^{2}\sum_{n=1}^{n_{a}}\sum_{m=1}^{n_{a}}\exp [-\frac{D_{\delta }(b_{mn})}{2}]
\end{eqnarray}
where $D_{\delta,atm}(b_{mn})$ is the variance of the random phase difference for the baseline $b_{mn}=|{\bf p}_{n}-{\bf p}_{m}|$:
\begin{equation}
D_{\delta,atm}(b_{mn})=<(\delta_{n,atm}-\delta_{m,atm})^{2}>
\end{equation}
It is assumed that the random phase difference $\Delta = \delta_{n,atm}-\delta_{m,atm}$ has normal probability distribution with zero mean and variance $\sigma_{\Delta}^{2}$.  In this case (7) was obtained using this relation:
%\begin{equation}
\begin{eqnarray}
<\exp (j\Delta )>=\frac{1}{\sqrt{2\pi }\sigma _{\Delta }}\int_{-\infty }^{\infty }\exp (j\Delta )\exp (-\Delta /2\sigma _{\Delta }^{2})d\Delta =\nonumber\\
\exp (-\sigma _{\Delta }^{2}/2)
\end{eqnarray}
%\end{equation}
Therefore, the loss  produced by the phase errors is equal to:
\begin{equation}
L_{\delta,atm }=\frac{P_{\delta \neq 0}({\bf u}_{s})}{P_{\delta =0}({\bf u}_{s})}=\frac{1}{n_{a}^{2}}\sum_{n=1}^{n_{a}}\sum_{m=1}^{n_{a}}\exp [-\frac{D_{\delta }(b_{mn})}{2}].
\end{equation}

The variance $D_{\delta,atm }(b_{mn})$ is the structure phase function of the turbulent atmosphere. The power-law (Kolmogorov) model will be used in the following sections to describe $D_{\delta ,atm}(b_{mn})$ both for the ionosphere and troposphere phase fluctuations (Tatarskii 1978).

\subsection{Ionosphere}
Phase fluctuations due to the irregular spatial distribution of the refraction index during wave propagation through the  ionosphere are described with the power-law model of the turbulence spectrum. The electron density $N$ in the ionosphere,  considered as a function of spatial coordinates, has variations which are characterized by a structure function of electron density $D_{N}(b)$ (Thompson et al. 2001, ch. 13):
\begin{equation}
D_{N}(b)=C^{2}_{N}b^{\gamma},
\end{equation}
where $\gamma= 2/3$, $C^{2}_{N}$ is the constant, $b$  is the baseline. This formula can be rewritten for the structure function of the refraction index:
\begin{equation}
D_{n}(b)=C^{2}_{n}b^{\gamma},
\end{equation}
where $C^{2}_{n}=\frac{r_{e}^{2}\lambda ^{4}}{4\pi ^{2}}C_{N}^{2},r_{e}=2.82\cdot 10^{-15}m$ (electron radius), $\lambda$ is the wavelength. Finally, the ionosphere phase structure function is:
\begin{equation}
D_{ion}(b)=2.91k^{2}C_{n}^{2}hb^{5/3}=2.91r_{e}^{2}\lambda ^{2}C_{N}^{2}hb^{5/3},
\end{equation}
where $h$ is the total propagation length through the irregularities of the ionosphere, $\lambda, h$ and $b$ must be substituted in meters.
The value $C^{2}_{N}$ can be estimated from (11) assuming that the ionosphere irregularities of electron density $\Delta(N)$ have a maximum dimension equal to $L_{0}$:

\begin{equation}
\Delta(N)^2=C_{N}^{2}L_{0}^{2/3}.
\end{equation}
For example,\\
for  $\Delta(N)/N=0.01$ and $N=10^{12} m^{-3}$ (day time) we have for $L_{0}=10km, C_{N}^{2}=2.154\cdot 10^{17}m^{-20/3}$,\\
 for $\Delta(N)/N=0.03$ and $L_{0}=30km, C_{N}^{2}=9.322\cdot 10^{17}m^{-20/3}$.
Figure 1a  shows the square root of the ionosphere structure function for $N=10^{12} m^{-3}$, $\Delta(N)/N=0.01$ and $L_{0}=10km, h=300km$ calculated for three frequencies: 50MHz, 100MHz and 200MHz.\\
\begin{figure}
\epsfig{figure=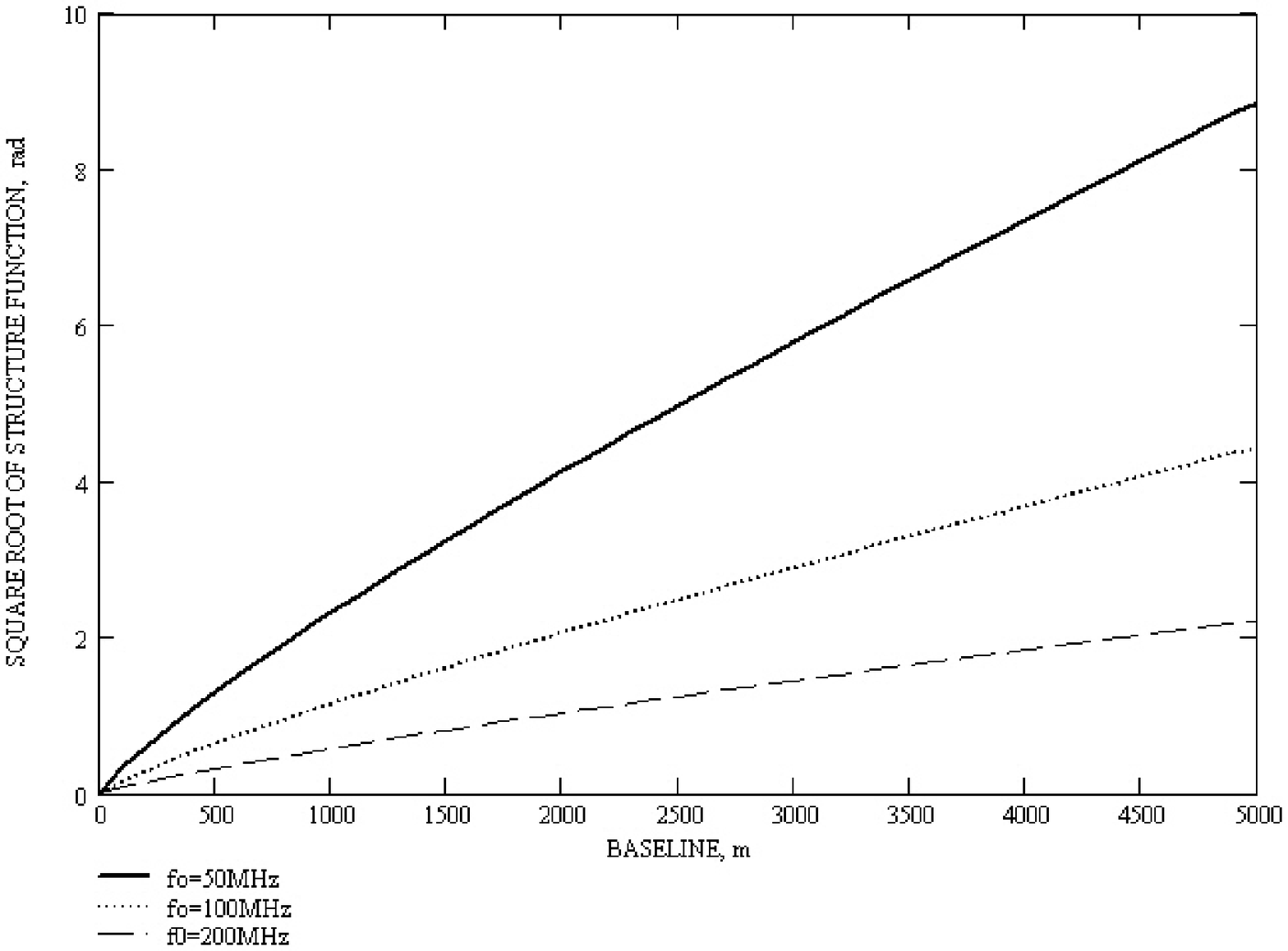,height=8.0cm,width=8.0cm}
\epsfig{figure=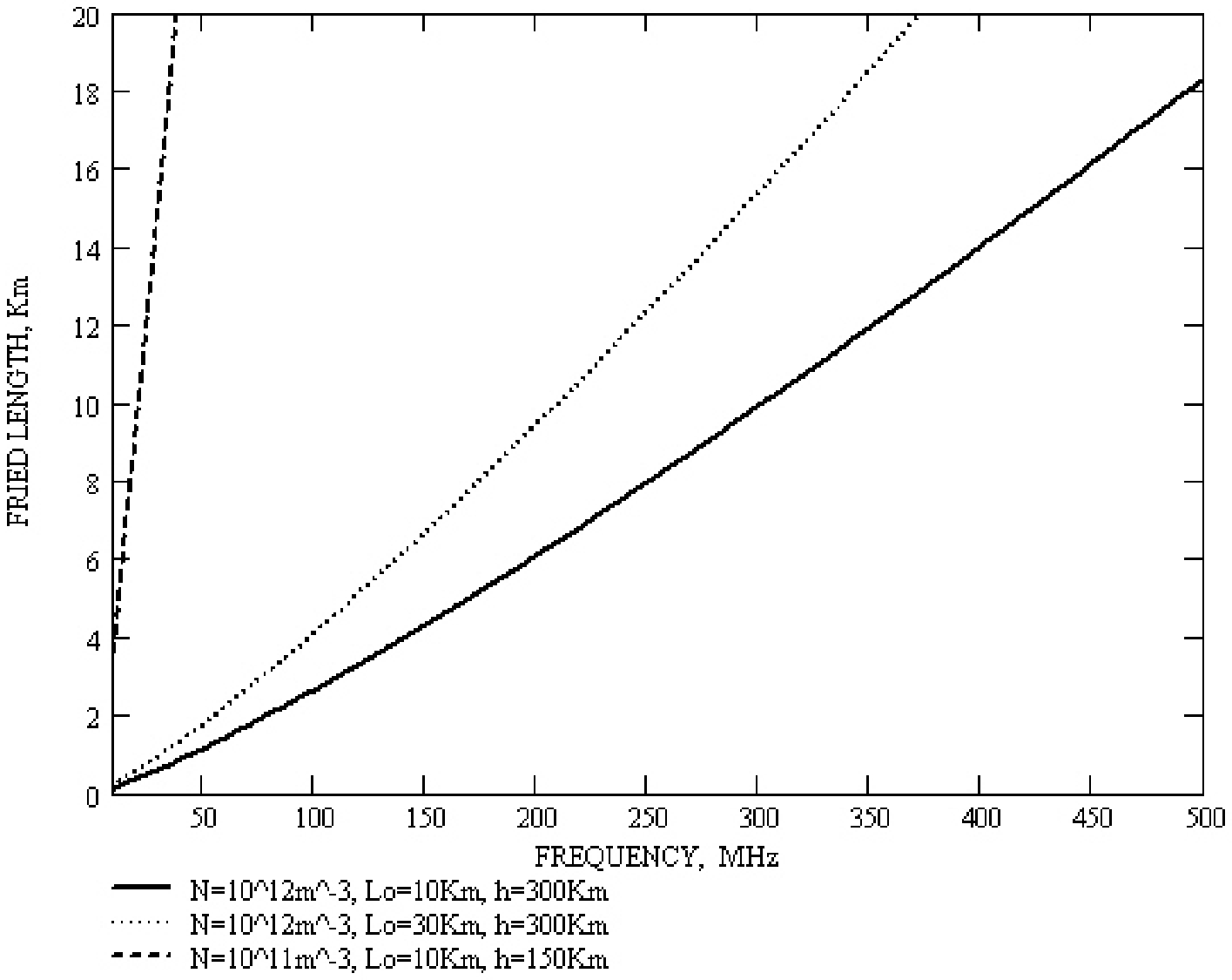,height=8.0cm,width=8.0cm}
\caption{a) upper panel: square root of the ionosphere structure function of electrical length, in cm; $N=10^{12} m^{-3}$ , $\Delta(N)/N=0.01$ and $L_{0}=10km$ calculated for three frequencies: 50MHz, 100MHz and 200MHz;
b) lower panel: Fried length for  different values of $N, L_{0},h$.}
\end{figure}
Phase fluctuations  can be also characterized  by the Fried length:
\begin{equation}
r_{Fried}=3.18d_{0},
\end{equation}
where $d_{0}$ is the baseline at which $\sqrt{D_{ion}}=1rad$. For the parameters used in  Fig. 1a  $r_{Fried}=1.16km$ for 50MHz, $r_{Fried}=2.66km$ for 100MHz and
$r_{Fried}=6.11km$ for 200MHz.  Fig. 1b  shows how Fried length depends on the frequency  for the different values  of $N, L_{0},h$.\\
The minimal time interval at which it is necessary to repeat calibrations can be calculated from Fried length: $t_{cal}=r_{Fried}/(3.18v_{wind})$ where
$v_{wind}$ is the wind velocity. Thus, for example, for  $r_{Fried}=6.11km$ and $v_{wind}=50m/sec$, $t_{cal}=38sec$.

Now the loss produced by ionosphere random phase errors  can be calculated  in the example of the array whose configuration is shown in Fig. 2a.
It is the random planar 100-element array with   the coordinates $x_{i}$ and $y_{i}$  represented by random normal values with zero mean and standard
deviation $SC$. For the array shown in Fig. 2a,  $SC=1000m$, therefore the maximum baseline is $\approx 5000m$.  The distribution of baselines (histogram) is shown in Fig. 2b   . Phase errors are maximal for the largest baselines but their relative number is less than the medium size baselines, therefore the signal loss for the array should take  this particular distribution of baselines into account.  Fig. 3   demonstrates the dependance of loss
versus  array size $5\cdot SC$. The curves are calculated  for three frequencies: 50, 100 and 200MHz.
\begin{figure}
\epsfig{figure=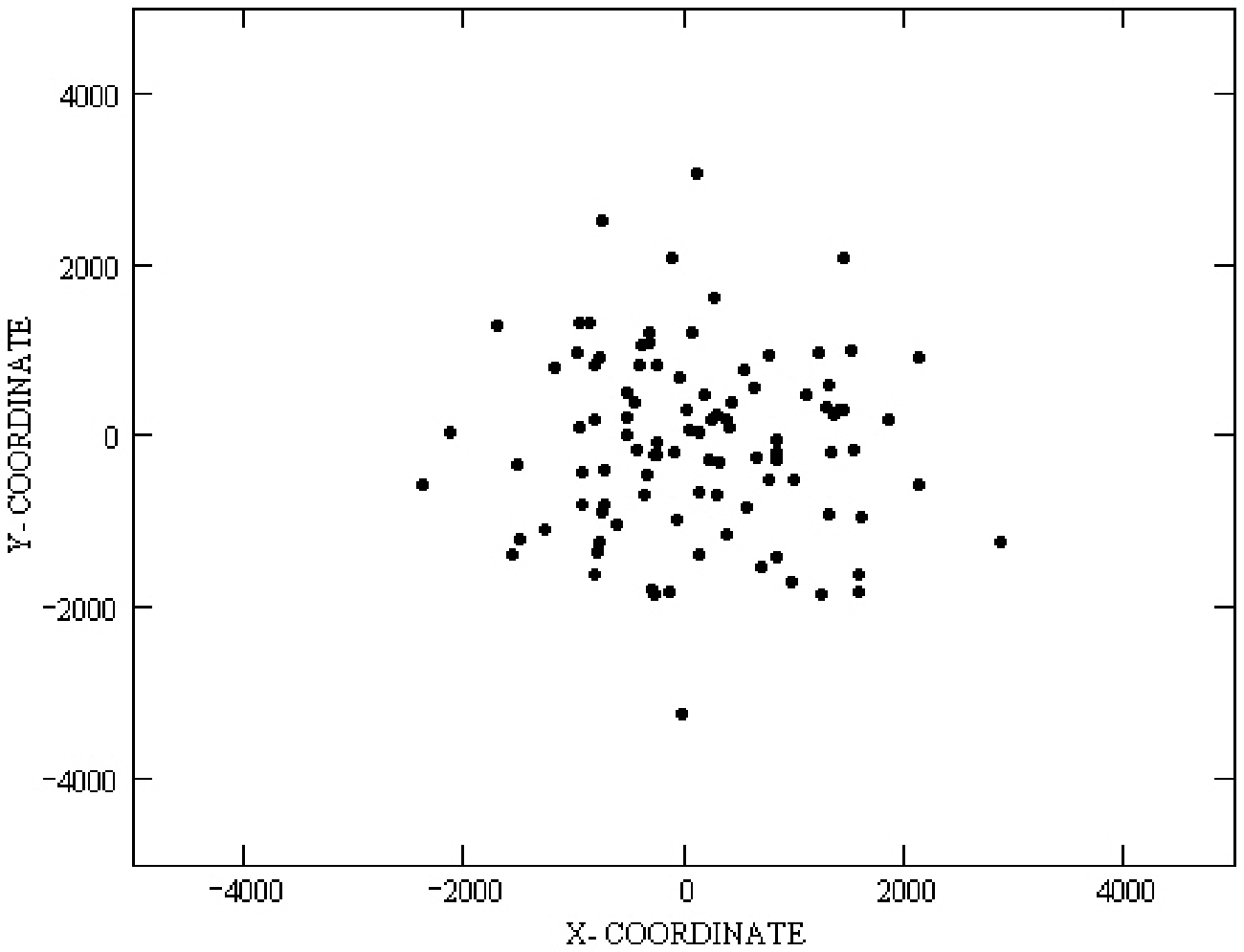,height=8.0cm,width=8.0cm}
\epsfig{figure=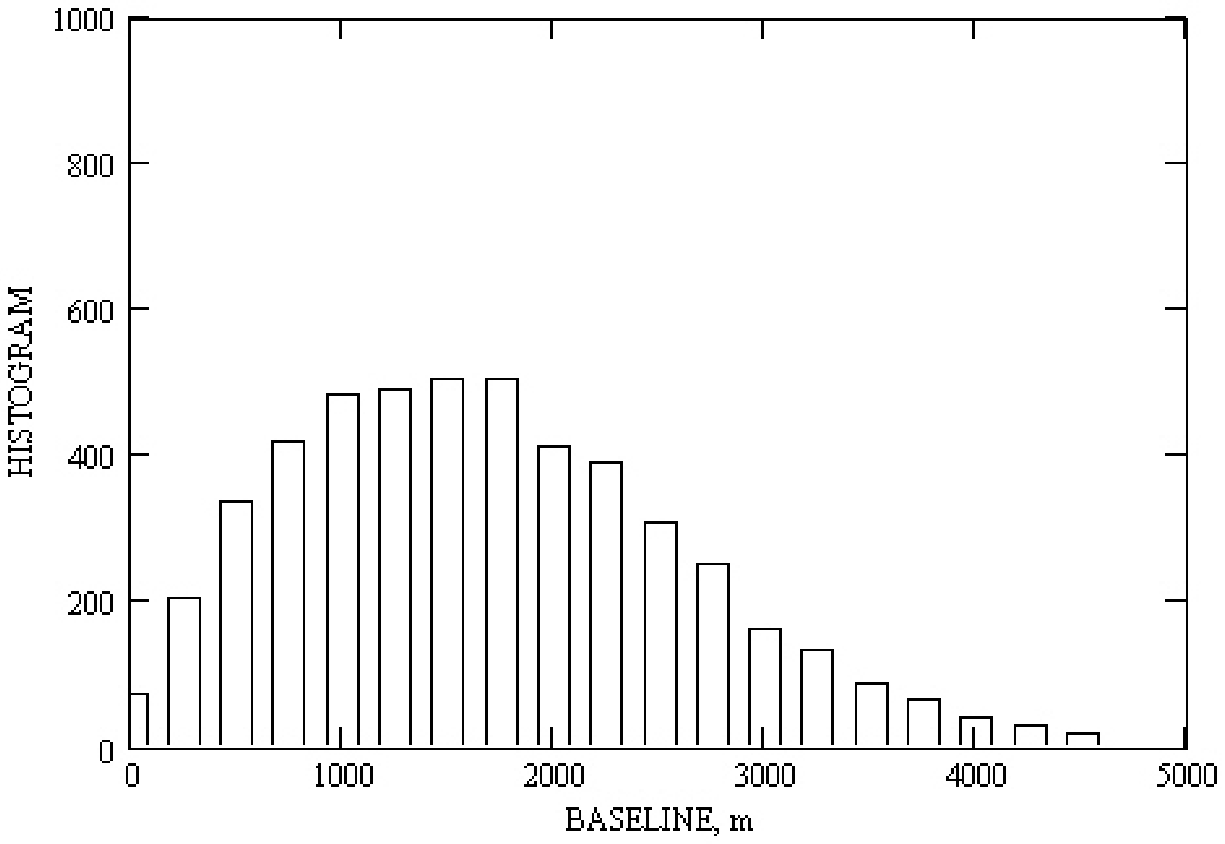,height=8.0cm,width=8.0cm}
\caption{a) upper panel: random array configuration; b) lower panel: histogram of the baselines.}
\end{figure}
\begin{figure}
\centerline{\epsfig{figure=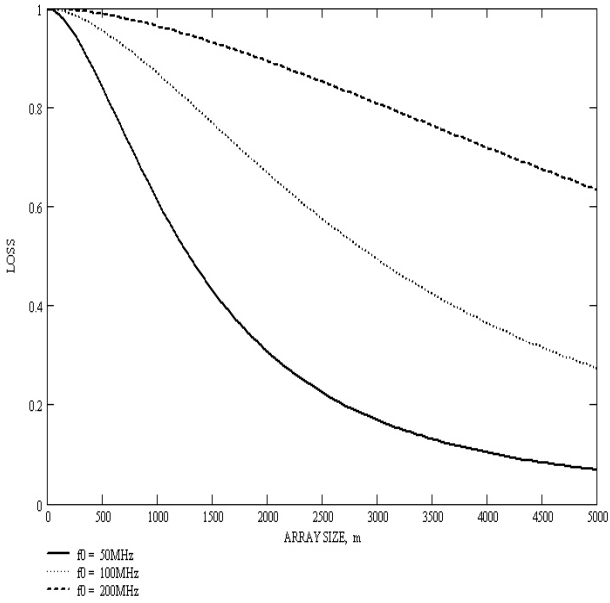,height=8.0cm,width=8.0cm}}
\caption{Loss produced by phase fluctuations in the ionosphere calculated for three frequencies: 50, 100 and 200MHz. The structure function from Fig. 1a is used. }
\end{figure}
\subsection{Troposphere}
Phase fluctuations due to the irregular spatial distribution of the refraction index during wave propagation through  the troposphere are also described in the frame of the power-law spectrum turbulence model. The troposphere  phase structure function is (Stotskii 1973, Carilli et al. 1999):
\begin{eqnarray}
D_{trop}(b)& = & 2.91k^{2}C_{l}^{2}b^{5/3},L_{0}<b<L_{1}\nonumber\\
& = &2.91k^{2}C_{L}^{2}b^{2/3}, L_{1}<b<L_{2}\nonumber\\
& = & 2.91k^{2}C_{L}, L_{2}<b,
\end{eqnarray}
where $L_{0}$ and $L_{1}$ are the internal and external  scales, respectively, of the isotropic three-dimensional turbulence model,
$L_{0}=0.1 -  1cm$, $L_{1}=5.6Km$ and $L_{2}=2000  - 3000Km$, the latter is determined by global meteorological variations.\\
Factors $C_{l}^{2}$ and $C_{L}^{2}$ depend on the local content of water vapor and oxygen in the troposphere (weather conditions)
 and the values chosen for the purpose of calculation are $C_{l}^{2} = 6.23 \cdot 10^{-11}m^{1/3}$ and $C_{L}^{2} = 3.64 \cdot 10^{-7}m^{4/3}$.\\
Fig. 4a  represents the structure function (16) and Fig. 4b  shows the   Fried length as a function of the frequency. Fig. 5   demonstrates the dependance of loss
versus  array size $5\cdot SC$. The curves are calculated  for three frequencies: 1400, 5000 and 8400MHz.
\begin{figure}
\epsfig{figure=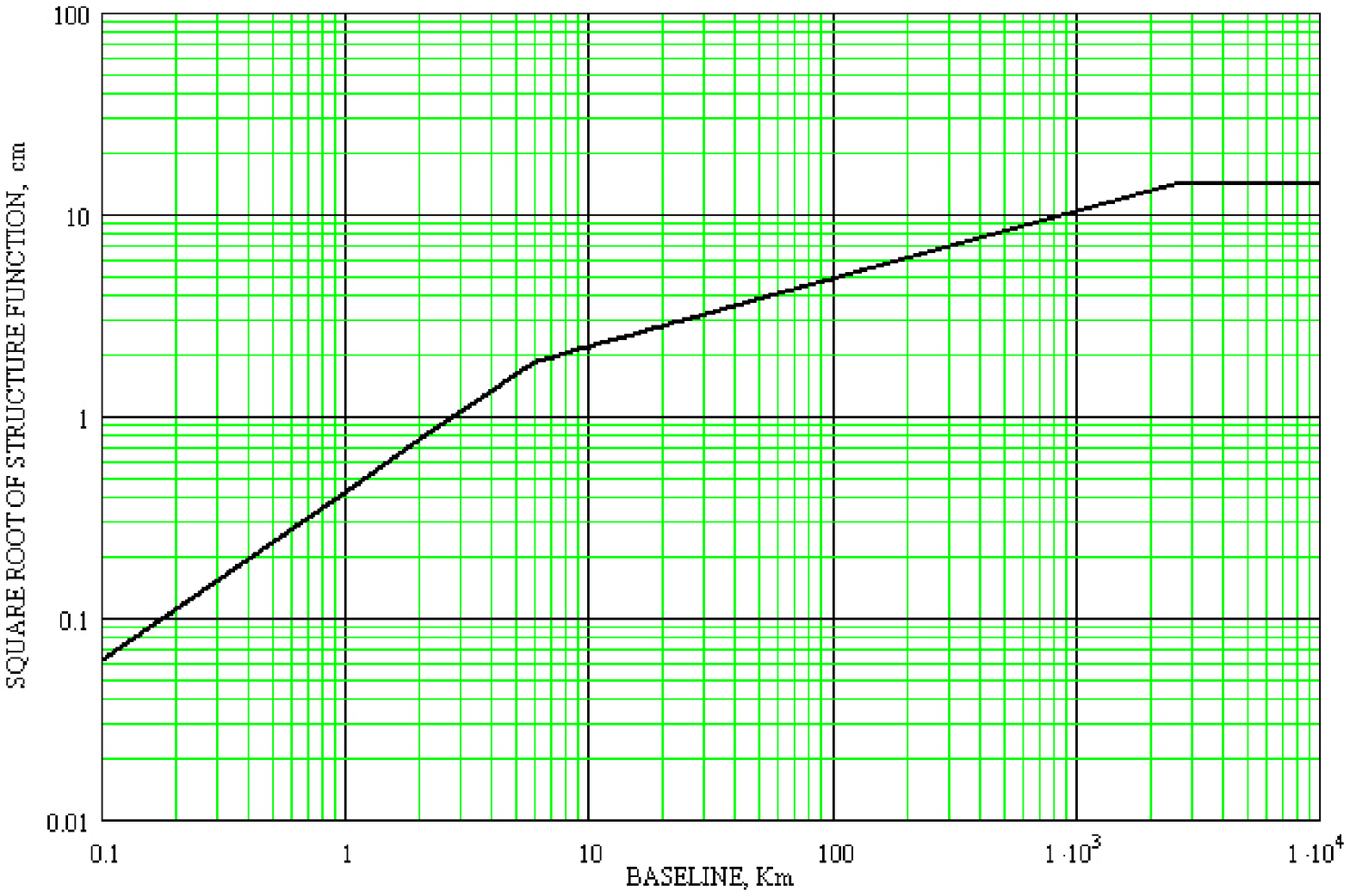,height=8.0cm,width=8.0cm}
\epsfig{figure=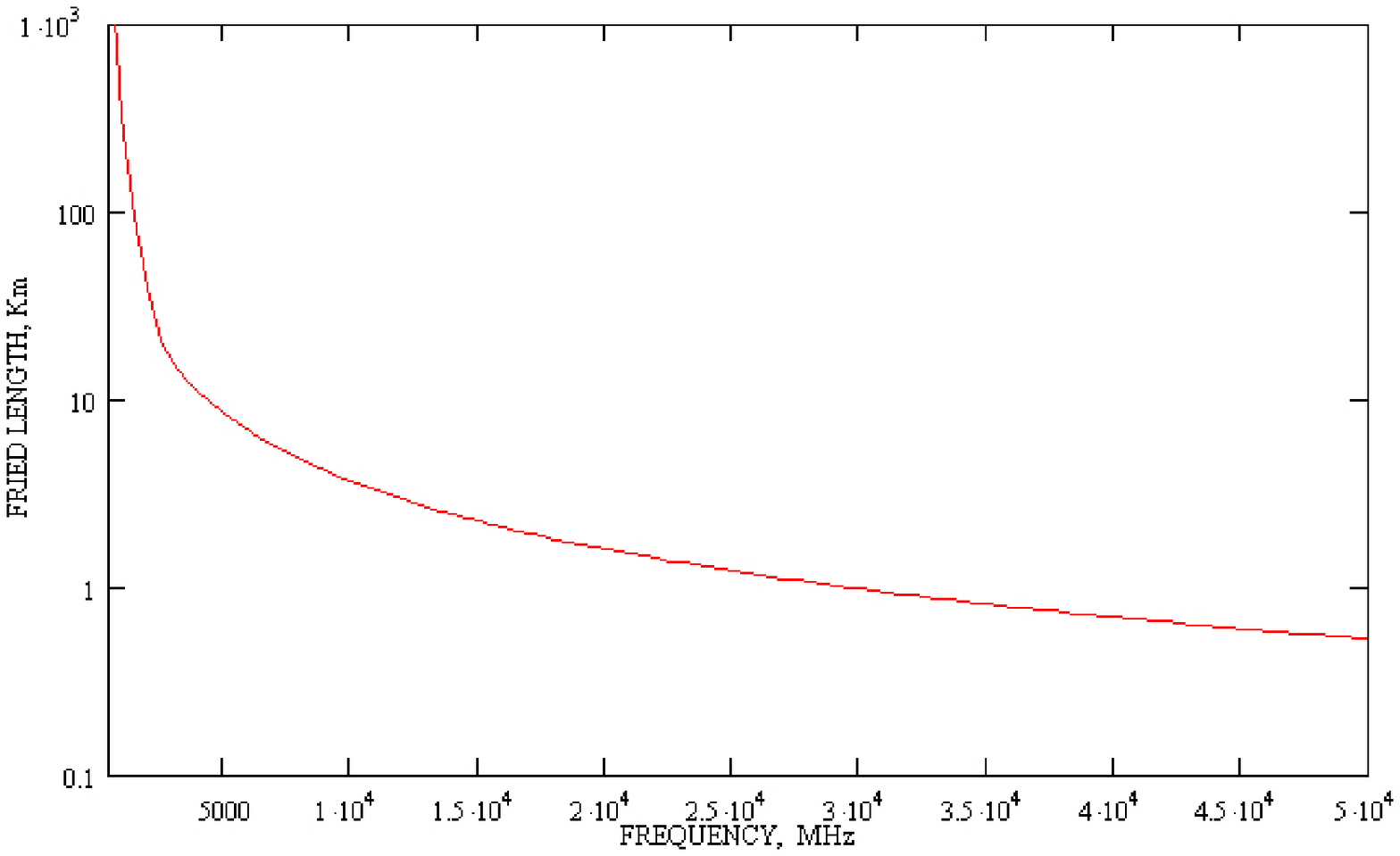,height=8.0cm,width=8.0cm}
\caption{a) upper panel: Square root of the troposphere structure function of electrical length, in cm;
 b) lower panel: Fried length as a function of frequency.}
\end{figure}
\begin{figure}
\centerline{\epsfig{figure=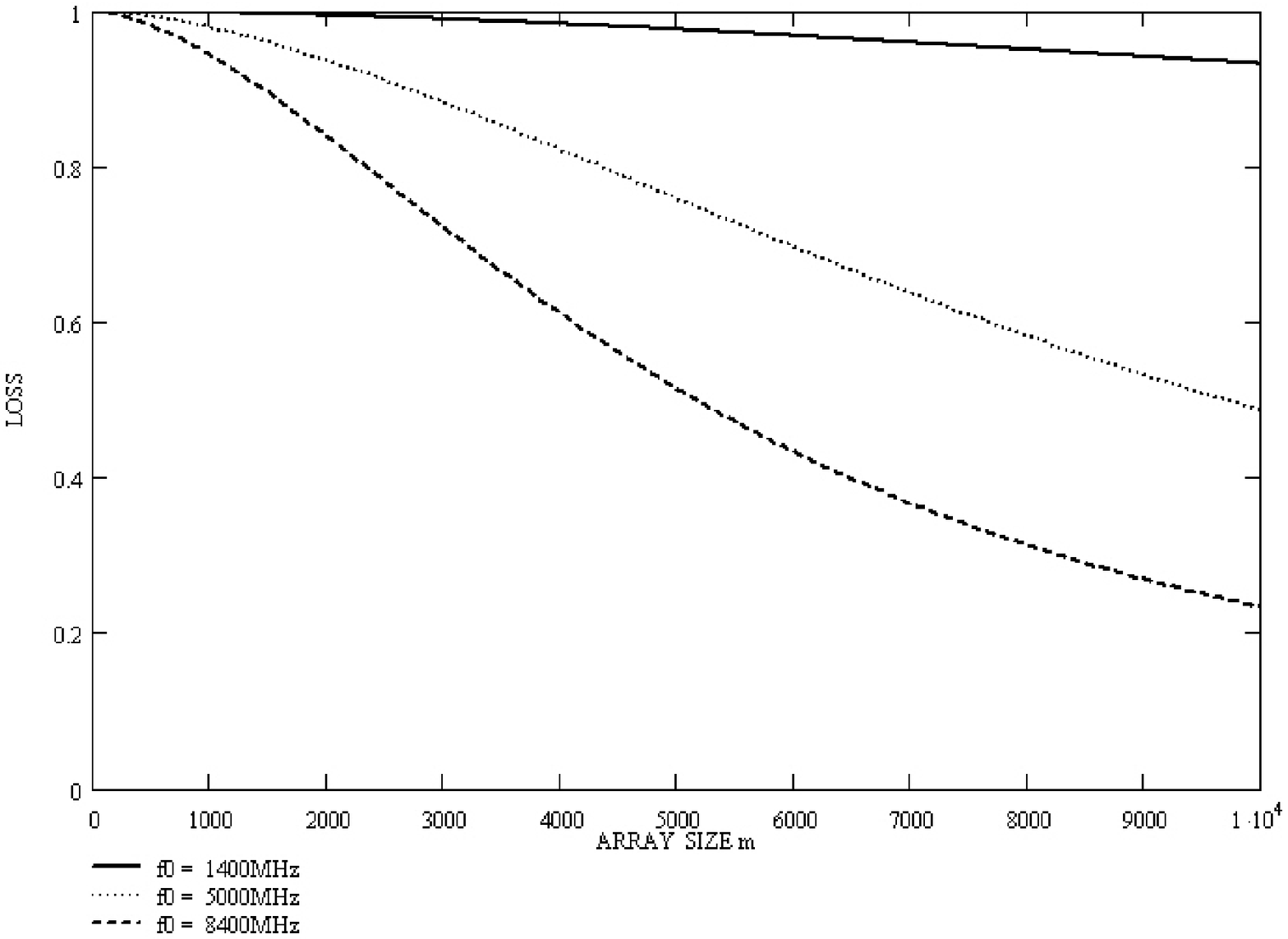,height=8.0cm,width=8.0cm}}
\caption{Loss produced by phase fluctuations in the troposphere calculated for three frequencies: 1400, 5000 and 8400MHz. The structure function from Fig. 4a is used. }
\end{figure}

\section{Self-cohering}
Observations in the tied-array mode (VLBI, transients, DtE) are pre-planned at any time and it is impossible to postpone them in order to choose better atmospheric conditions (for example, night time in the case of the ionosphere).  The effectiveness of the synthesized aperture must be maximal during observations which means that periodical calibrations are necessary. Traditional methods of radio interferometer calibration can be applied using the grid of calibration point sources.
This  calibration must be made in {\it real time} with the help of available correlators which must work in parallel with the tied-array adder.
Here, another method is proposed which uses the points in the direct images of the field-of-view with calibration sources.

 It is presumed that a full calibration  with the correlator has already been performed  before  each tied-array observation.
During  subsequent observations  the total power output of the tied-array is used as a tracking tool and the proposed algorithm
will introduce small  phase corrections at  short time intervals therefore keeping  the amplitude of the calibration source at a prescribed level.
This has similarities  to the approach of
(Muller\&Buffington 1974) which is also described in
(Tyson 1991).

The choice of calibration sources is the same as in  traditional  methods.

Equation (7) corresponds to the synthesized beam when there are  phase errors $\delta _{n,atm}$ produced during  propagation through the turbulent atmosphere.
 To eliminate $\delta _{n,atm}$,  compensation phase shifts $\delta _{n, comp}$
are introduced at each $n$-th array element.
The phase of the signal corresponding to the direction ${\bf u}_{s}$ at the n-th array element is
\begin{equation}
 \theta _{n}({\bf u}_{s})=(2\pi /\lambda )(-{\bf p}_{n}{\bf u}_{s})+\delta _{n}+\delta _{n,atm}-\delta _{n,comp},
\end{equation}
where $\delta _{n,comp}$ is the compensation phase shift introduced  for the correction of $\delta _{n,atm}$ - atmospheric phase error.
The value   of the signal power in the prescribed direction ${\bf u}_s$  in the  image with one or several calibration sources available in the field-of-view (FoV) can be obtained by convolution of the sky intensity $B_{0}({\bf u})$ with the synthesized beam $P_{comp}({\bf u}_{s})$
\begin{equation}
B_{comp}({\bf u}_{s})=B_{0}({\bf u})\oplus P_{comp}({\bf u}_{s}),
\end{equation}
where $\oplus$ denotes convolution.
The similar approach to imaging  procedure with an array  is considered in  the {\it direct imaging} scheme, (Wright 2004).
Having the output of the total power detector  $B_{comp}({\bf u}_{s})$ for the direction ${\bf u}_{s}$ we can maximize this value by varying  $\delta _{n,comp}$, i.e., we have an optimization problem for $n_{a}$ values of compensation phases. In general, for each direction ${\bf u}_{s}$ this problem can be written as
\begin{equation}
\widehat{\delta }_{n,comp}({\bf u}_{s})=\arg \max_{\delta _{n,comp}}[B_{comp}({\bf u}_{s})],n=1...n_{a}
\end{equation}

This  scheme is illustrated in the following example of computer simulation. The 60-element spatially random planar array is represented in Fig.  6. The phase errors introduced in each of the array element signals are modeled by the two-dimensional random  value, Fig. 7, with circular symmetrical  spatial spectral density which decreases radially  according to the (-11/3) power law. The instantaneous sample of phase errors at frequency 100MHz as a function of baseline length is given in Fig. 8. This high level of  phase errors has been chosen to demonstrate the effectiveness of the proposed method.\\
The image containing three point sources is represented in Fig. 9 (left panel) and the synthesized image in the presence of the phase errors (Fig. 8) is shown in Fig. 9 (middle panel).
(isoplanicity being presumed).

The value of the synthesized image in the direction of the largest source (lower left in the image) was used as the cost function. The genetic algorithm(GA) was applied because of the strong multi-modality of the cost function (19) and this algorithm finds the global maximum successfully.
 Genetic algorithms  search for  a solution to a set of variables by the use of simulated evolution, i.e., the survival of the fittest strategy.
In contrast to  calculus-based algorithms (conjugate gradients and quasi-Newtonian methods), GA, first introduced in (Holland 1975), exploit a guided random
technique during optimization procedure (Goldberg 1989, Michalewicz 1994, Charbonneau 1995).\\
GA optimizers are particularly effective when the goal is to find an approximate global maximum in a high-dimension, multi-modal function domain in a near-optimal manner.
They are also largely independent of the starting point or initial guess.
There is  parallelism which allows for the exploitation of several areas of the solution space at the same time. This parallelism can be very useful in the implementation of GA on the multi-core platform and FPGA.
In this article, computer simulation has been done on a PC (Intel Pentium, 2.5 Ghz, 1GB RAM) using Matlab 7.6.0. Specific GA operations (selection, crossover and mutation) have taken approximately $3\%$ of the
total computing time: 150 sec for 100 iterations (each iteration is the full cycle of these operations). The rest of the computing time was spent on the calculation of the cost function
(formation of the beam with corrected phases  $->$ convolution with the image $->$ total power output). But these calculations are necessary only in computer simulations:
in reality the values of the cost function are supplied  by the tied-array itself (``Nature'' does the job).

After applying the optimization procedure   and introducing the resulting compensation phases the corrected image is shown in Fig. 9 (right panel).\\
The contour  presentations in Fig. 10  correspond to the undistorted image (left panel), the image with phase errors (middle panel) and the image after correction (right panel), respectively.\\
The corresponding synthesized beam is restored up to 0.94 of its undistorted value.

There are some peculiarities in image processing with non-planar arrays
(Perley 1999) but the tied-array mode concerns
only point-like sources. Therefore, there is no difference in planar and non-planar arrays in the context of this article (phase irregularities due to  atmospheric turbulence), especially for the adaptive
calibration procedure described here.

\begin{figure}
\epsfig{figure=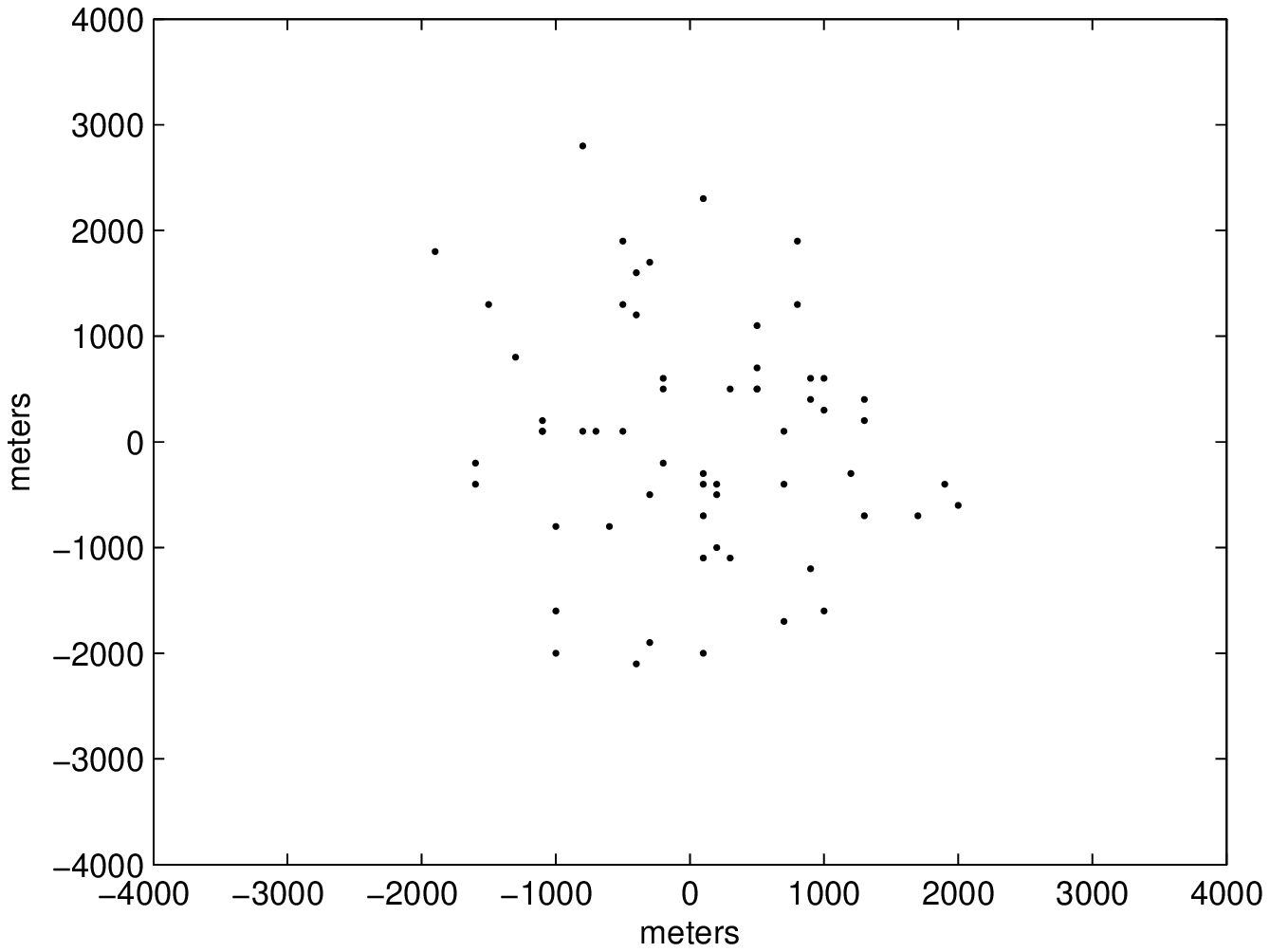,height=6.0cm,width=8.0cm}
\caption{60-element tied array configuration.}
\epsfig{figure=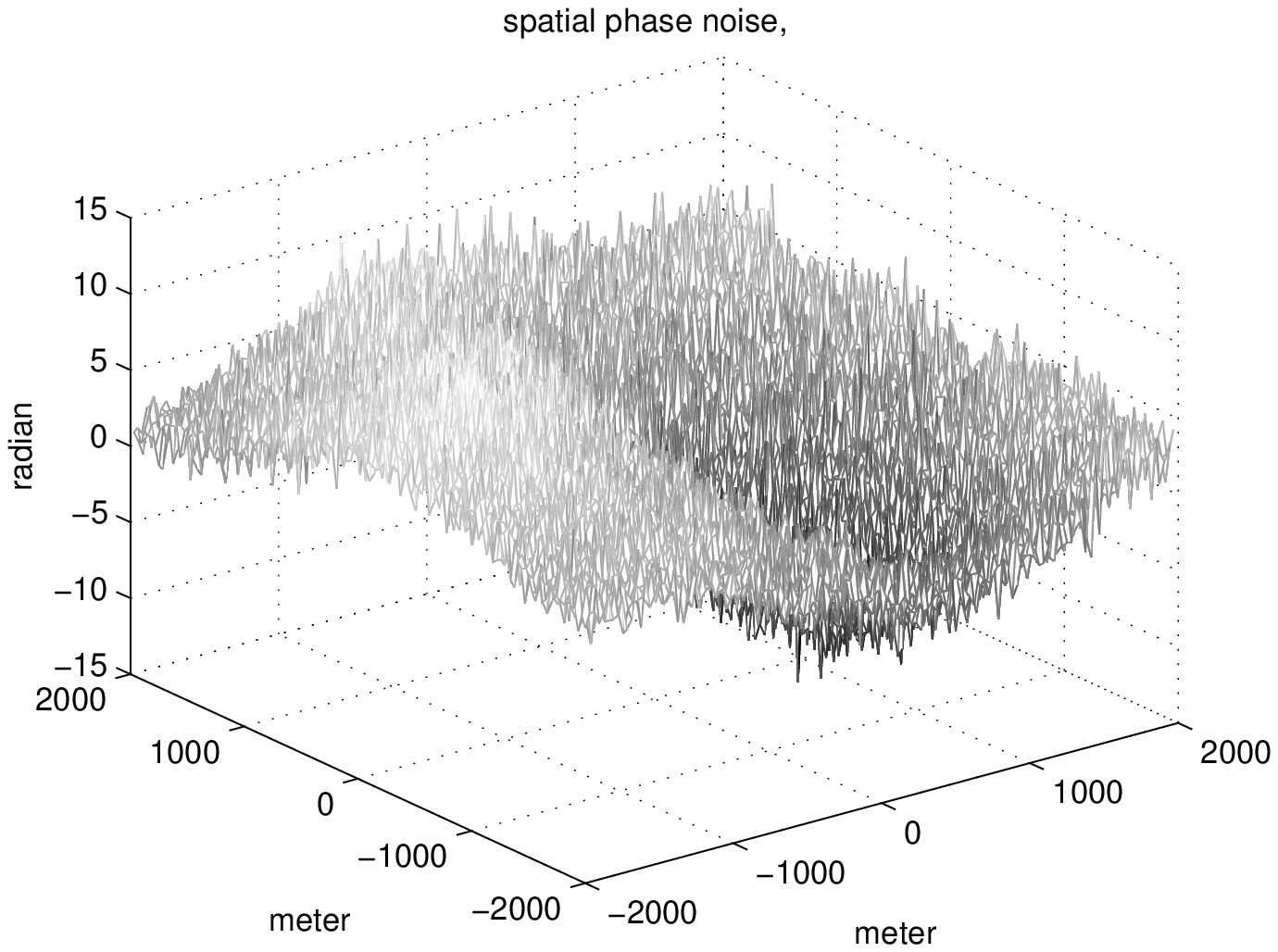,height=6.0cm,width=8.0cm}
\caption{Spatial phase error  distribution, projected on the array plane.}
\epsfig{figure=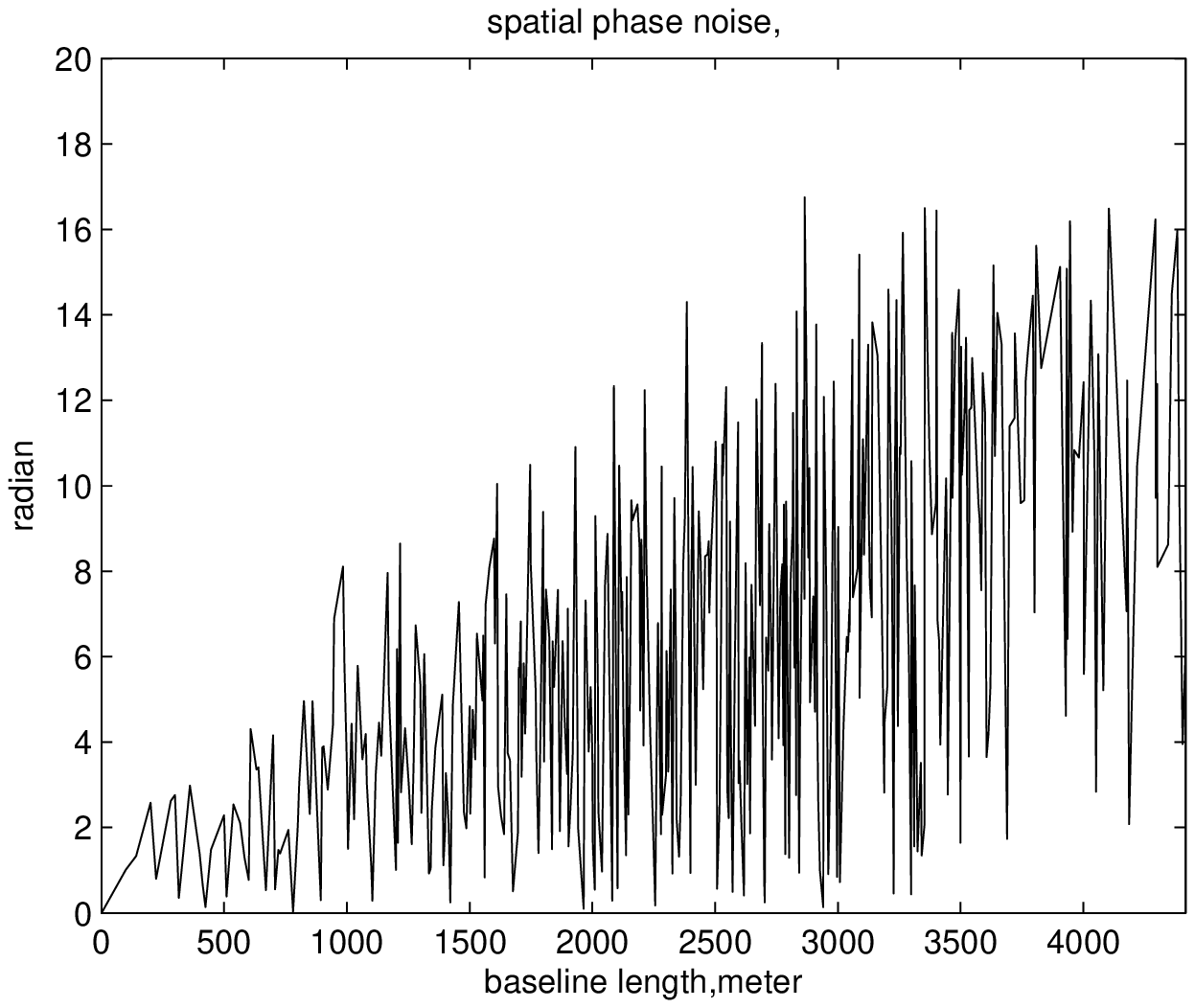,height=6.0cm,width=8.0cm}
\caption{Phase errors as function of baseline length.}
\end{figure}

\section{Conclusions}
\begin{enumerate}
  \item The effective area of tied arrays may be  significantly reduced  by ionospheric and tropospheric phase irregularities at low and high frequencies, respectively.
  \item Observations  are made at times (VLBI, transients monitoring, DtE)  when it is impossible to choose   quiet atmospheric conditions and  real-time calibration is necessary and has to be fulfilled  in parallel with  observations.
 \item  The total power  at the  auxiliary outputs of the tied-array, phased in the direction of calibration sources,
can be  used on a level  with traditional  calibration methods using correlators.  Multi-beam facilities are necessary for creating these  auxiliary outputs.
Optimization algorithms (genetic algorithms, simulated annealing) can be used to compensate for propagation phase errors by maximizing the amplitude of a chosen calibration  source.
 The tied array can  preserve its correctly phased state during lengthy observations using one or several auxiliary outputs, therefore working
in the self-cohering regime.
The proposed scheme does not exclude  traditional methods of calibration - it is complementary to them.
\end{enumerate}

\begin{acknowledgements}
    I am grateful to Roy Smits whose comments were very
     helpful.
\end{acknowledgements}

\onecolumn
%\nopagebreak
\begin{figure}
\epsfig{figure=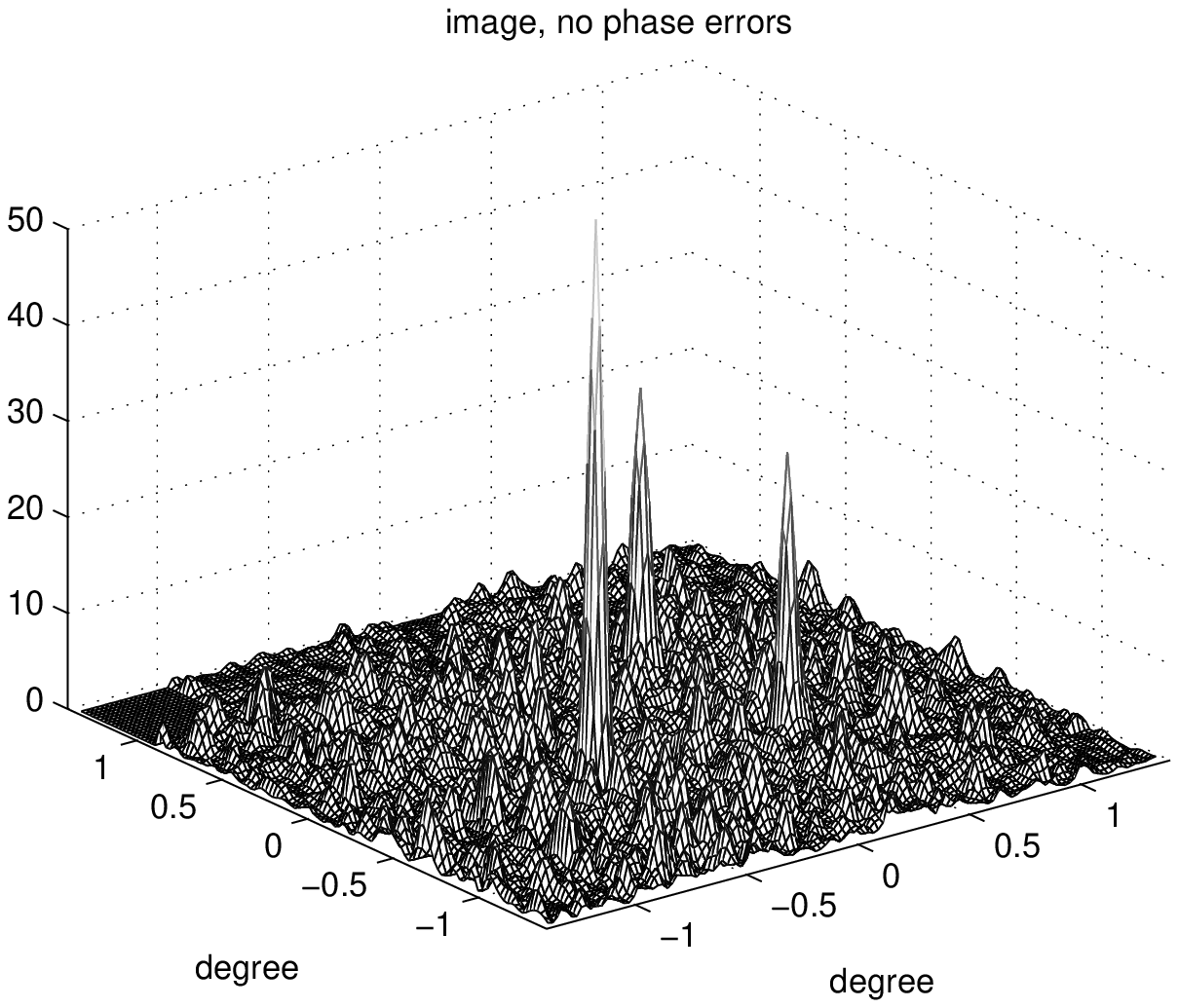,height=6.0cm,width=6.0cm}
%\caption{The synthesized image without phase errors.}
\epsfig{figure=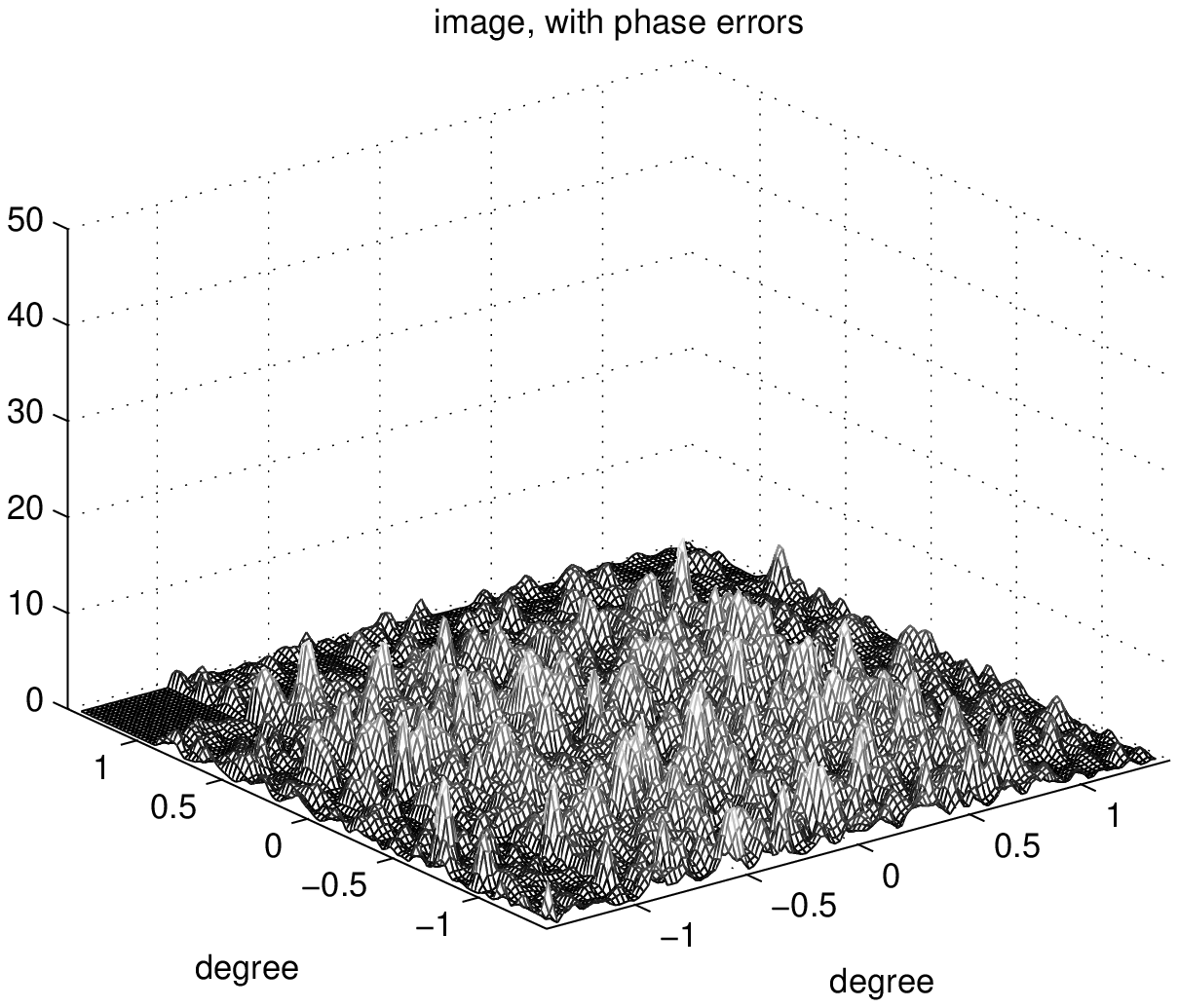,height=6.0cm,width=6.0cm}
%\caption{The synthesized image with phase errors.}
\epsfig{figure=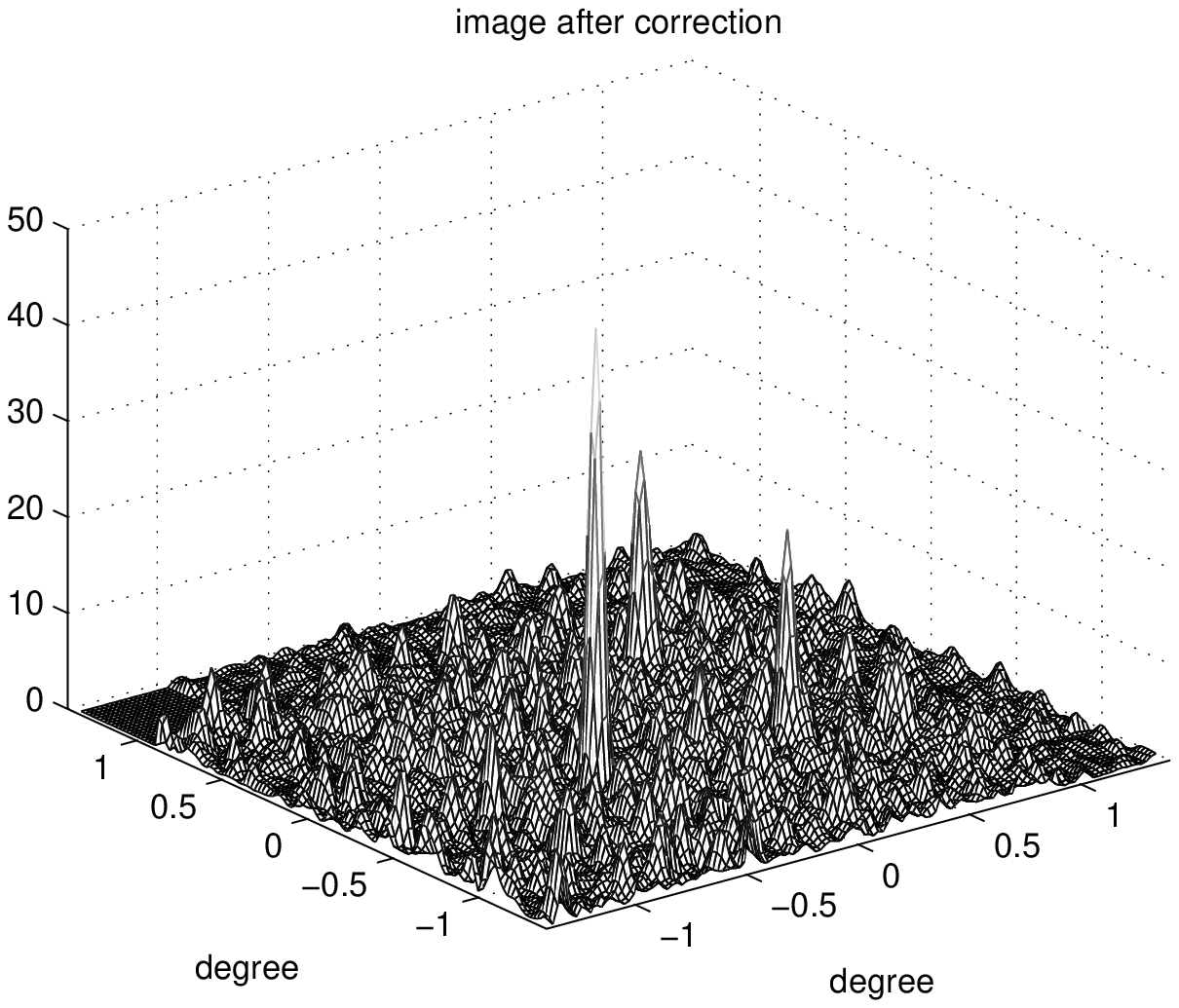,height=6.0cm,width=6.0cm}  %
\caption{Left panel:  synthesized image without phase errors;
middle panel: synthesized image with phase errors;
right panel: synthesized image after correction.}
\end{figure}

\begin{figure}
\epsfig{figure=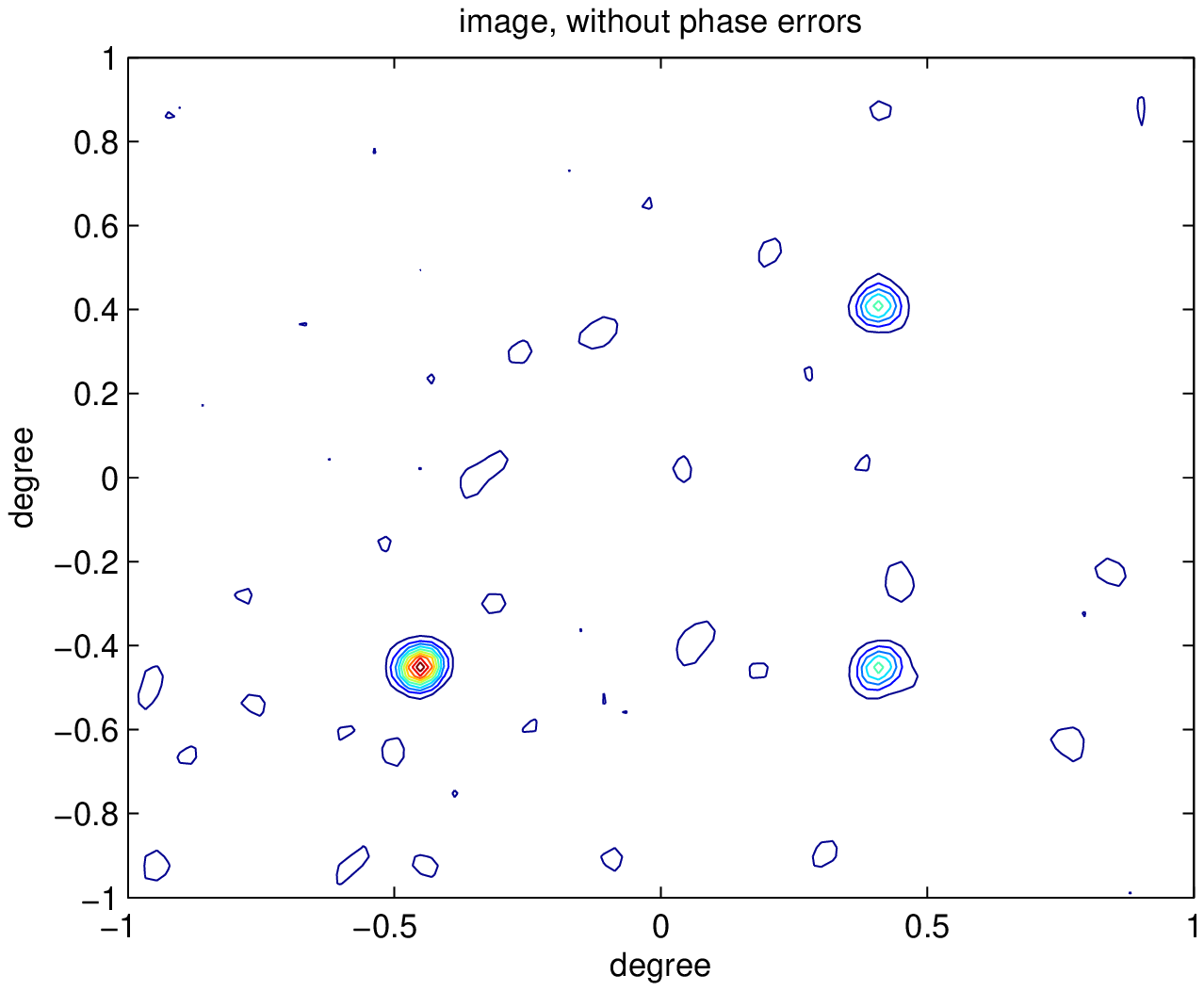,height=6.0cm,width=6.0cm}
%\caption{Contour presentation of the synthesized image without phase errors.}
\epsfig{figure=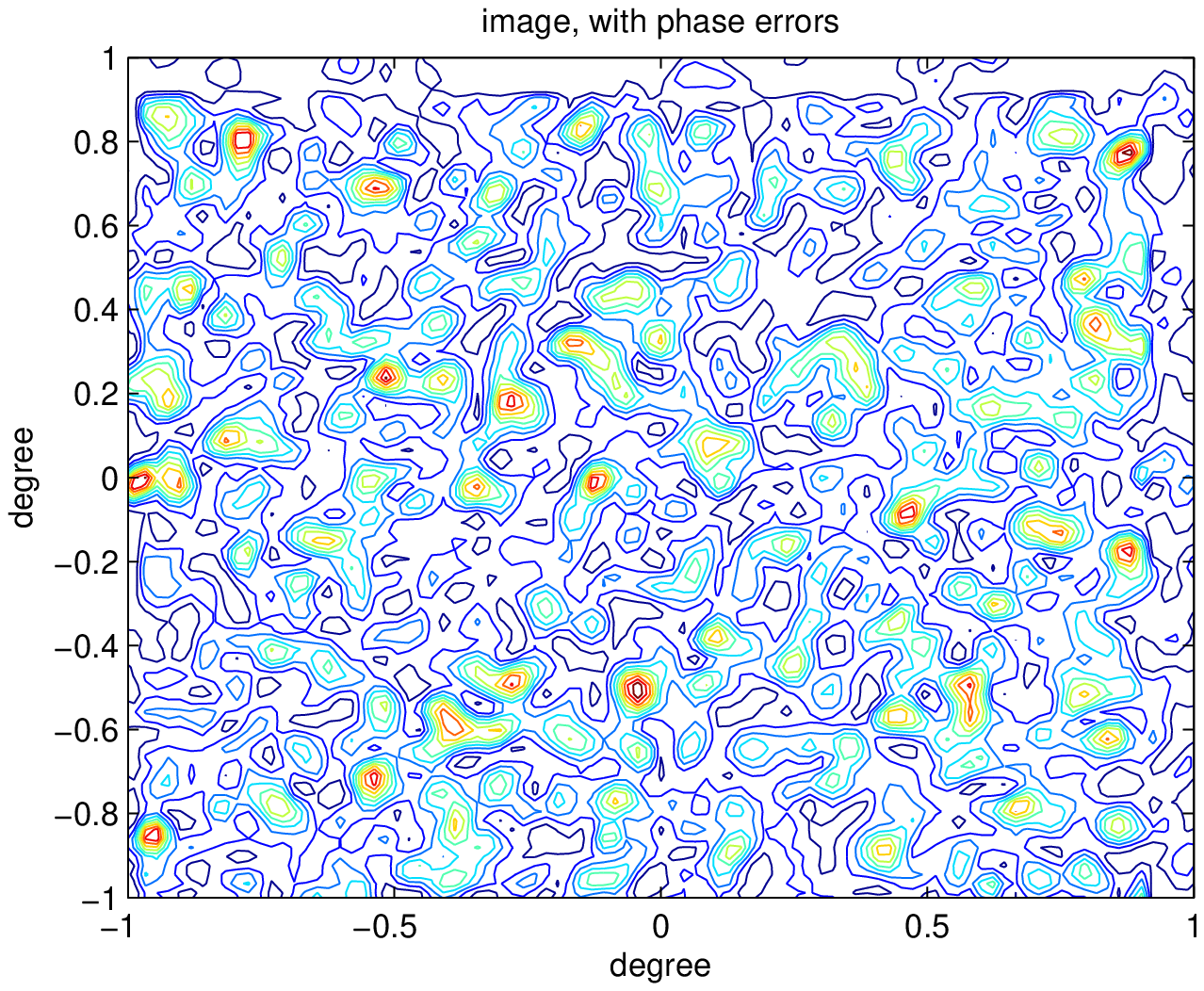,height=6.0cm,width=6.0cm}
%\caption{Contour presentation of the synthesized image with phase errors.}
\epsfig{figure=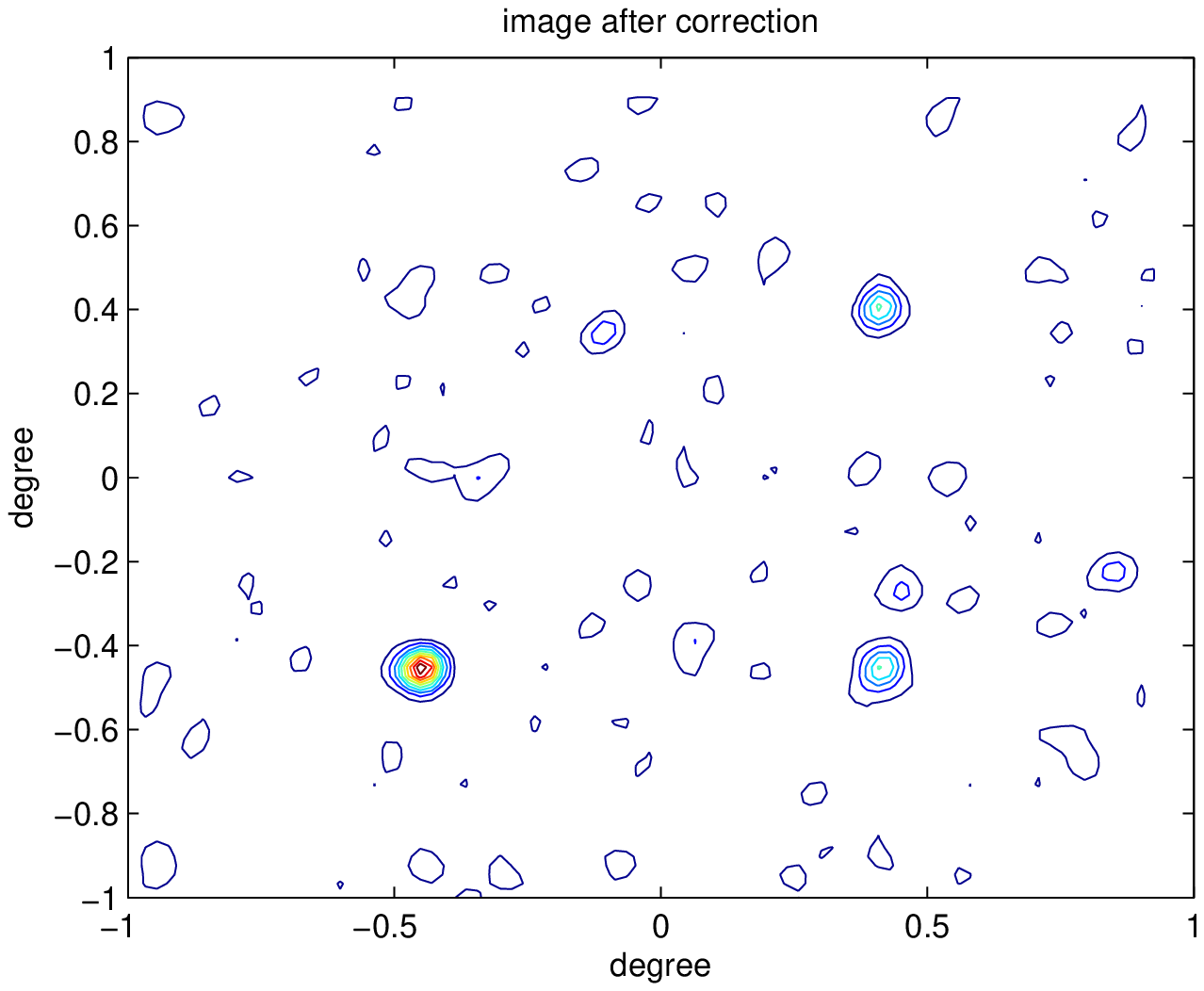,height=6.0cm,width=6.0cm}
\caption{Contour presentations, left panel: synthesized image without phase errors;
middle panel: synthesized image with phase errors; right panel:  synthesized image after correction.}
\end{figure}
\nopagebreak

\end{document}